\documentclass[trackchanges,twocolumn]{aastex701}
\hypersetup{linkcolor=blue,citecolor=blue,filecolor=blue,urlcolor=blue}
\usepackage{parskip}
\usepackage{hyperref}

\begin{document}

\title{The Research Impact of the Jodrell Bank Observatory and Other Facilities affected by the UK Science Funding Cuts in 2025}

\author[orcid=0009-0004-2970-6805]{Rommulus Francis Lewis}
\affiliation{Department of Physics, University of Hong Kong, Hong Kong, Hong Kong SAR}
\affiliation{Hong Kong Institute for Astronomy and Astrophysics, University of Hong Kong, Hong Kong, Hong Kong SAR}
\email[show]{rommulus@connect.hku.hk}  

\author[orcid=0000-0003-1276-1248]{Amruth Alfred}
\affiliation{Department of Physics, University of Hong Kong, Hong Kong, Hong Kong SAR}
\affiliation{Hong Kong Institute for Astronomy and Astrophysics, University of Hong Kong, Hong Kong, Hong Kong SAR}
\email[hide]{rommulus@connect.hku.hk}

\author[orcid=0009-0007-8431-1322]{Hetansh Shah} 
\affiliation{Department of Computer Science, University of Massachusetts, 300 Massachusetts Ave, Amherst, MA 01003, United States}
\email[hide]{rommulus@connect.hku.hk}

\collaboration{all}{The Astrophysics Wrapped Collaboration}

\begin{abstract}

The United Kingdom Research and Innovation body and the Science and Technology Facilities Council recently announced funding cuts to facilities at the Jodrell Bank Observatory, specifically the e-MERLIN network of radio telescopes as well as a few other facilities across the world like the James Clerk Maxwell Telescope. These funding cuts have been extremely disturbing to the Astronomical community, as is evident from widespread news coverage about these cuts. Here we present a short analysis of the research impact e-MERLIN and the other affected facilities had in 2025 from a dataset consisting of every Astrophysics paper submitted to the arXiv during the year. This serves to help members of the community and all stakeholders make better sense of the research output of the facilities that are not being prioritised under the new budget.

\end{abstract}


\section{Introduction} 
In January 2026, the Science and Technology Facilities Council (STFC), part of the United Kingdom Research and Innovation (UKRI), the body responsible for managing funding for scientific research in the United Kingdom (UK), announced funding cuts to Particle Physics, Astronomy, and Nuclear Physics, collectively known as PPAN, on the order of 30\%. The agency also warned scientific project leaders to consider the possibility of cuts extending to a drastic 60\% of the 2024-2025 budget\footnote{\href{https://www.ras.ac.uk/news-and-press/news/proposed-budget-cuts-catastrophe-uk-astronomy}{RAS Article on Initial Funding Cut}}. These cuts would reduce funding for PhD students and early career researchers and would threaten the UK's ability to contribute to major international collaborations. Considering such drastic cuts, the scientific community was naturally unhappy. In response, a group of science communicators, physicists and astronomers from the UK, including Professor Brian Cox, Professor Jim Wild, Dr Rebecca Smethurst and more, met with Members of Parliament (MPs) in June to discuss and initiate change to the planned funding cuts (for more detailed information please read the numerous online resources such as the RAS articles on this issue\footnote{\href{https://ras.ac.uk/news-and-press/news/mass-cuts-and-potential-jodrell-closure-devastating-astronomy}{RAS Article on New Funding Cuts}\label{footnote_ras}}).

As a result of this meeting and public pressure, the STFC announced that the funding cuts would be limited to 2.7\% and post-doctoral researchers would be protected. While this was a step in the right direction, it came at a cost. According to the new prioritisation scheme, the James Clerk Maxwell Telescope (JCMT) would no longer be funded and funding for the Birmingham Solar Oscillations Network (BiSON), the Extremely Large Telescope (ELT), the Square Kilometre Array (SKA) and the Vera C Rubin Observatory (Rubin) would be partially reduced\footref{footnote_ras}$^{,}$\footnote{\href{https://www.bbc.com/news/articles/c5y6r9rdxygo}{BBC Article on Funding Cuts}}.

Jodrell Bank Observatory, part of the Jodrell Bank Center for Astrophysics at the University of Manchester, has held significant historical value ever since its inception in 1945 by radio astronomer Bernard Lovell. The biggest instrument at the observatory, the Lovell Telescope, was constructed with gun turret mechanisms from World War 1 battleships and was the world's largest steerable radio telescope at the time of construction and is the third largest today. The telescope was instrumental in the tracking of Soviet and US satellites and spacecraft during the Space Race and, in many cases, was the only telescope capable of doing so. In 1960, the launch of the Pioneer 5 probe was crucially dependent on the Lovell Telescope sending commands to the probe and was the only telescope in the world capable of receiving data from the probe. Amongst many scientific discoveries at the Jodrell Bank Observatory, some striking examples are the discovery of the first gravitational lens in 1979 by Jodrell Bank astronomer Dennis Walsh, and the discovery of the first Einstein ring with e-MERLIN - an interferometric array with telescopes across England with its base at Jodrell Bank Observatory. This is only a brief overview meant to lay the historical background for the reader, while the focus of this manuscript is to examine the recent scientific, educational and societal contributions of the instruments affected by the impending funding cuts.

\section{Methods}
To quantify the impact of each telescope on astrophysical research in 2025 we search through the LaTeX source files for all Astrophysics papers submitted to the arXiv\footnote{\url{https://arxiv.org}} (a hub for researchers in the field to release their work as preprints before publication) in 2025. We use string matching such that if a particular paper mentions the names of any of the telescopes from a pre-defined list, that paper is counted as using the data from that telescope. For example, when considering e-MERLIN, we search through the paper's LaTeX for any mention of the words `e-MERLIN', `Lovell Telescope', `Knockin Telescope', etc, including any of the other telescopes in the network. We also allow for case-insensitive searching for the long forms of these telescopes and case-sensitive searching for the short forms. Since e-MERLIN is also an important part of the European Very Long Baseline Interferometer (EVN) network, we also include (as a separate analysis) how its contribution changes when EVN is included as a search term. We follow a similar analysis for the other affected facilities. For a more detailed explanation of the extraction, please refer to \cite{wrapped2025}. 

\section{Analysis}
We split our analysis into two parts, one where we only consider the e-MERLIN network since that has been the focus of the media in these funding cuts. The second part focuses on the other facilities affected by the proposed cuts. 

\subsection{e-MERLIN}
\begin{figure}
    \centering
    \includegraphics[width=\linewidth]{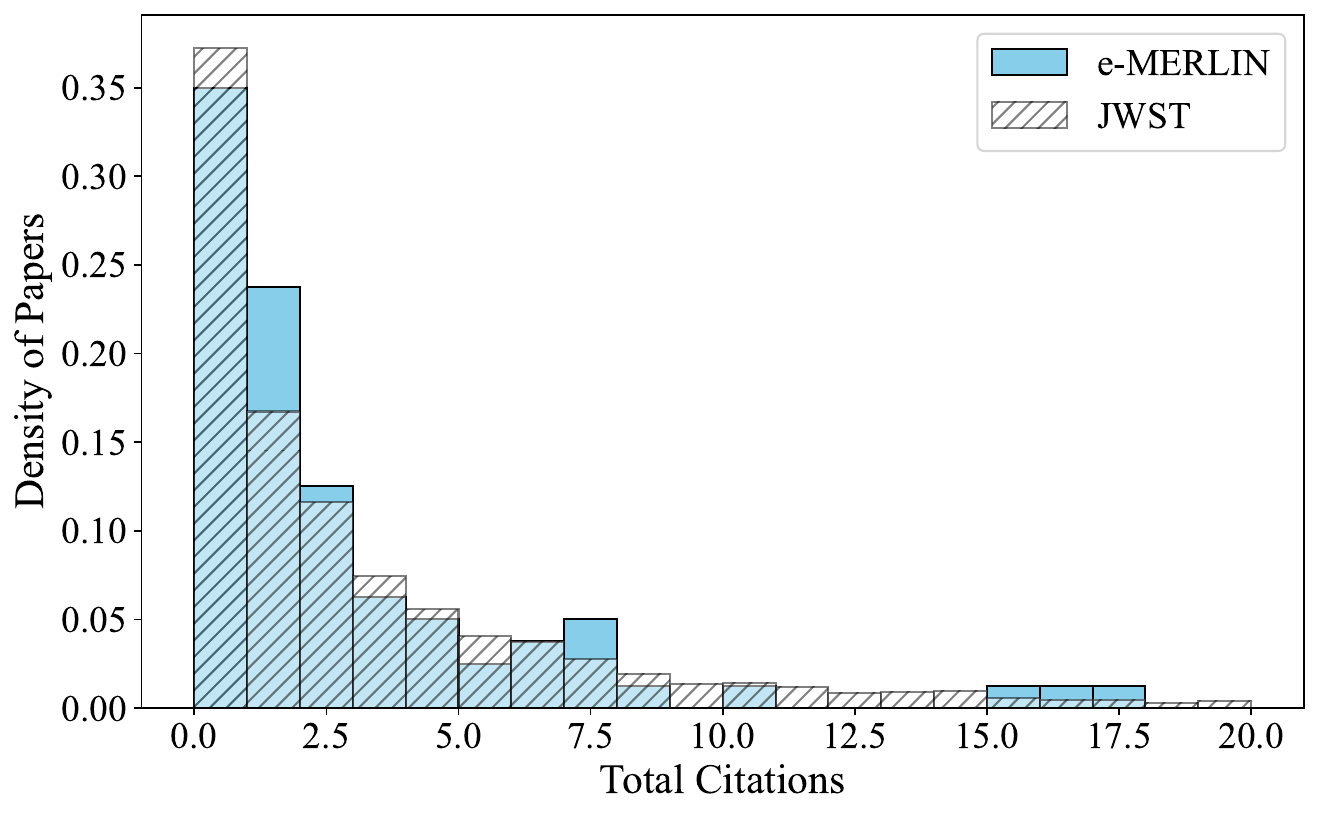}
    \caption{\textbf{Normalised Distribution of All Papers Mentioning JWST and e-MERLIN:} This plot demonstrates that the average number of citations is not skewed by an excess of high or low citation papers, as the distribution follows very closely the citation distribution of a very highly cited telescope like JWST \citep{wrapped2025}.}
    \label{fig:1}
\end{figure}

When searching only for terms relating to `e-MERLIN' and the individual telescopes part of the network, we find a total of 80 papers in 2025 mentioned using data from the e-MERLIN network. In total, all these papers were cited 192 times (direct citations indexed as of the last week of December 2025). On average, a paper using data from e-MERLIN was cited 2.40. For context, papers that mention JWST are cited an average of 3.93 times per paper. Figure \ref{fig:1} shows the normalised distribution of citations compared with the James Webb Space Telescope (JWST) to ensure the average is not skewed because of outliers. One reason for this could be the fact that e-MERLIN is nowadays used to obtain important follow-up observations of high-energy astrophysical phenomena such as X-ray transients discovered with the Einstein Probe (EP) as is the case of \cite{yadav2025radioobservationspointmoderately} (which is also the most cited paper that used e-MERLIN) as well as for Gamma Ray Bursts (GRB) like \cite{anderson2025radioflaremultiwavelengthafterglow}. These high-energy astrophysical phenomena are a fairly new branch of the field, with important discoveries and progress being made every day, which is why such papers are often highly cited. We note that the top two primary subjects under which papers mentioning e-MERLIN are submitted are `High Energy Astrophysical Phenomena' followed by `Astrophysics of Galaxies', which supports our previous argument. 

When we include terms relating to EVN in our search, we find the number of papers doubles to 161, as it allows papers like \cite{zhang2025flaringradiocounterpartfast} which only mention the EVN and not e-MERLIN specifically to be included in the count. Thus, e-MERLIN contributes an equal amount to astronomical research both individually and as a part of the EVN. With EVN included, the total number of citations extends to 384, working out to an average of 2.39 citations per paper. 

\subsection{JCMT, BiSON, ELT, SKA and Rubin}
While most of the news has been centred around the Jodrell Bank Observatory, several other facilities will also face budget cuts, and hence we present some statistics on these facilities as well.

The UK has completely withdrawn from the JCMT, which was mentioned in 235 papers in 2025, accumulating 402 total citations with an average citation rate of 1.71, with the top primary subject for JCMT papers being `Astrophysics of Galaxies'. 

For the three other instruments involved, the proposed funding cuts are partial. BiSON was attributed in 18 papers with 51 citations or an average of 2.83 citations per paper, with `Solar and Stellar Astrophysics' being expectedly the top primary subject. ELT was cited 1336 times with mentions across 490 papers, amounting to an average of 2.73 citations per paper. For the SKA (which is still under construction), there were 1416 papers mentioning the facility with 4520 citations, equivalent to an average of 3.19 citations per paper. Finally, the Vera Rubin Observatory was already the 9th most mentioned telescope in 2025 papers when only considering mentions in the title and abstract \citep{wrapped2025}. When including all text, there are 1874 papers quoting the Rubin Observatory with 7296 total citations and 3.89 citations per paper. Importantly, the Rubin Observatory only began operating halfway through 2025, meaning a good number of these papers were released before the first test images were taken.

\section{Conclusions}
We have provided selected statistics on the research output of all the facilities affected by the recently proposed science funding cuts in the UK, with a focus on the e-MERLIN interferometric array facility of the Jodrell Bank Observatory. We note that this analysis is subject to the same caveats outlined in \cite{wrapped2025}. Additionally, readers should note that the methodology to search for terms in this paper differs slightly from that mentioned in \cite{wrapped2025}; the difference being that here we search through the entire manuscript whereas in \cite{wrapped2025} we only search through the title and abstract of each paper. 

Being students and early career researchers, we differ to the judgment of the more experienced professionals in the field who’s years of experience and wealth of knowledge makes them better equipped to weigh both the research (from this work) and cultural impact of these instruments to asses if the funding cuts are justified or not. We acknowledge that given the budget assigned, something needs to be cut somewhere. We also remind readers that a 20\% cut (as an example) in funding need not directly translate to a 20\% reduction in research output. We hope readers, be it the general public, members of the community or policy-makers, will use these statistics to decide for themselves if these funding cuts are reasonable or not. Ultimately, we all want the best for science given the available budgets.

\begin{acknowledgments}
R.F.L and A.A would like to thank our supervisor at the University of Hong Kong, Jeremy Lim, for allowing us the freedom to take up projects like this which might not contribute to our main research interests but are extremely important for the community. All data used in this paper comes from papers publicly available on the arXiv. The citation information was indexed in the third week of December 2025 using the NASA ADS API, which in turn uses the Astrophysics Data System, funded by NASA under Cooperative Agreement 80NSSC25M7105.
\end{acknowledgments}

\bibliographystyle{aasjournalv7}
\bibliography{sample701}

\end{document}